\documentclass{jpsj-suppl}
\usepackage{times}
\usepackage{color}
\usepackage{bm}

\title{Spin-Spin Correlation Enhanced by Impurities in a Frustrated Two-leg Spin Ladder}

\author{Takanori Sugimoto$^{1,2}$\thanks{E-mail address: sugimoto.takanori@jaea.go.jp},
Michiyasu Mori$^{1,2}$,
Takami Tohyama$^{3}$,
and Sadamichi Maekawa$^{1,2}$
}
\inst{$^{1}$Advanced Science Research Center, Japan Atomic Energy Agency, Tokai, Ibaraki 319-1195, Japan \\
$^{2}$CREST, Japan Science and Technology Agency, Chiyoda-ku, Tokyo 102-0075, Japan \\
$^{3}$Yukawa Institute for Theoretical Physics, Kyoto University, Kyoto 606-8502, Japan
}
\abst{We theoretically study a spin-spin correlation enhanced by non-magnetic impurities in a frustrated two-leg spin ladder.  
The frustration is introduced by the next-nearest-neighbor antiferromagnetic interaction in the leg direction in the antiferromagnetic two-leg spin ladder.
The spin-spin correlation function around impurity site is calculated by the density-matrix renormalization-group method.
We find that the spin-spin correlation is enhanced around impurity site with the wavenumber reflecting the frustration.
As increasing the frustration, the wavenumber is shifted from commensurate to incommensurate. 
We discuss several experimental results on BiCu$_2$PO$_6$ in the light of our theory.}

\kword{Two-leg spin ladder, frustration, impurity, BiCu$_2$PO$_6$, Bi(Cu$_{1-x}$Zn$_x$)$_2$PO$_6$}

\begin{document}
\maketitle

\section{Introduction}
The effects of impurities on low-dimensional quantum spin systems have been intensively studied in relation to a spin gap, a long-range order and the spin-Peierls transition.
In general, antiferromagnetic (AF) quantum spin systems have the singlet ground state with and without a spin gap, where magnetization as a classical order parameter vanishes and the spin-spin correlation is not long-ranged.
In such a singlet ground state, several preceding works showed that non-magnetic impurities induce the magnetization in both the gapless and gapped systems.
Although long-range order emerges due to weak three dimensionality in real compounds, the magnetic order is strongly suppressed by the spin fluctuation in a low dimension.
Theoretically, impurity-induced magnetization can be understood by the freezing of unpaired spins in the singlet ground state.
For example, the AF spin-1/2 Heisenberg chain Sr$_2$CuO$_3$, does not have magnetization~\cite{kerren93,ami95,eggert96}. 
The AF magnetization around impurities, which are introduced by interstitial excess oxygen, has been observed by nuclear magnetic resonance (NMR) measurement~\cite{takigawa96}.
The magnetization can be interpreted as local spin freezing near the ends of the spin chains. 
In addition, the studies on the spin-Peierls system showed that non-magnetic impurity also induces a gapless dispersive-mode, which indicates that a long-range order emerges.
The long-ranged order induced by impurities is observed in the spin-Peierls system CuGeO$_3$~\cite{hase93}.
The spin dimerization and the bond alternation accompanied by a spin gap emerge below the critical temperature in the system.
The doped impurities break the spin dimers and generate free spins.
The free spins compose the AF long-range order with the Goldstone mode with the help of the lattice degree of freedom~\cite{fukuyama96,saito97,martin97,hirota98,gehring00}.

For the two-leg spin-ladder system SrCu$_2$O$_3$~\cite{azuma94}, where a spin gap exists and a spin-spin correlation decays exponentially, the effects of impurity have been investigated both theoretically and experimentally~\cite{azuma97,fujiwara98,ohsugi99}.
Doped impurities substituted Zn or Ni for Cu enhance the spin-spin correlation length~\cite{ohsugi99}.
In the system, spin configurations are dominated by the singlet dimers on the rung bonds.
The doped impurities break the rung singlets, and free spins are effectively generated against the rung-singlet background.
The free spins can correlate each other through a boson-like elementary particle, so-called {\it triplon} in the two-leg ladder system.

For a newly synthesized two-leg spin ladder BiCu$_2$PO$_6$, the NMR and the inelastic neutron scattering (INS) study reported that this compound has a strong frustration~\cite{koteswararao07,mentre09,tsirlin10}.
The effective spin model corresponds to the two-leg spin ladder with the next-nearest-neighbor AF interaction in the leg direction.
Thus, the compound is an interesting system bridging between SrCu$_2$O$_3$ and CuGeO$_3$.
Equivalently, the NMR and the muon spin rotation measurement on the Zn-doped BiCu$_2$PO$_6$ showed that impurity-induced magnetization emerges around impurity sites~\cite{bobroff09,wang10,casola10,alexander10,andrade12,lavarelo13}.
Surprising is that the spin-spin correlation of the ground state has an incommensurate wavenumber~\cite{lavarelo12,sugimoto13,plumb13}.
This is different from the non-frustrated two-leg spin ladder SrCu$_2$O$_3$ witn the commensurate wavenumber.
The incommensurate wavenumber will reflect the spin freezing caused by doped impurities.
That is, the magnetic order induced by impurities could be like a spiral order.
However, it has not been clear whether the spin-spin correlation enhanced by impurities reflects the incommensurability of the host system or not.

The rest of the paper is organized as follows. First, we present the model Hamiltonian of the frustrated two-leg spin ladder with non-magnetic impurities in Sec. II.
In this section, we define the spin-spin correlation functions calculated by using density-matrix renormalization-group (DMRG) method.  
In Sec. III, we demonstrate enhancement of the spin-spin correlation around impurities.
We also present the spin-spin correlation between two impurities and its wavenumber.
We discuss the spin-spin correlation from the viewpoint of frustration. 
In Sec. IV, we summarize the present results and discuss recent experimental data in the light of our calculations.

\section{Model and Method}
\begin{figure}
\begin{center}
\includegraphics[scale=0.5]{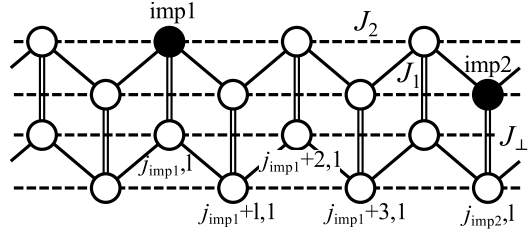}
\end{center}
\caption{A frustrated two-leg spin ladder with two impurities.  Open circles represent spin-$\frac{1}{2}$ sites.  Impurities are denoted by closed circles.  The nearest- and next-nearest-neighbor interactions in the leg direction, $J_1$ and $J_2$ are marked with solid and dashed lines, respectively. A double solid line denotes a nearest-neighbor interaction in the rung direction $J_\perp$. Two impurities are located in the upper chain and the position of sites in the lower chain is denoted by the distance from the impurities like $j_\mathrm{imp1}+2, \mathrm{l}$.}
\label{fig:tlzz_imp}
\end{figure}

The model Hamiltonian of the frustrated two-leg spin ladder in Fig.~\ref{fig:tlzz_imp} is given by,
\begin{equation}
\mathcal{H}=\mathcal{H}_{1}+\mathcal{H}_{2}+\mathcal{H}_{\perp},
\end{equation}
with
\begin{eqnarray}
\mathcal{H}_{1}&=&J_1\sum_{j,i=\mathrm{u,l}} \bm{S}_{j,i} \cdot \bm{S}_{j+1,i}, \\
\mathcal{H}_{2}&=&J_2\sum_{j,i=\mathrm{u,l}} \bm{S}_{j,i} \cdot \bm{S}_{j+2,i}, \\
\mathcal{H}_{\perp}&=&J_\perp\sum_j \bm{S}_{j,u} \cdot \bm{S}_{j,l}.
\end{eqnarray}
Here, $J_1 (>0)$ and $J_2 (>0)$ are the magnitudes of the AF nearest-neighbor and next-nearest-neighbor exchange interactions, respectively, and $J_\perp (>0)$ is that of the AF nearest-neighbor interaction in the rung direction. 
$\bm{S}_{j,\mathrm{u(l)}}$ is the $S=1/2$ spin operator on the $j$ site in the upper (lower) chain.

We introduce impurity as $\bm{S}_{\mathrm{imp}}=\alpha \, \bm{\sigma}/2$ on an impurity site where $\alpha$ is the positive value and $\bm{\sigma}$ is the Pauli matrix.  
In the limit $\alpha\to 0$, the site with $\alpha$ is associated with a non-magnetic impurity.
We introduce two impurities in the upper chain as shown in Fig.~\ref{fig:tlzz_imp}.

To investigate impurity effects, we use two types of spin-spin correlation functions,
\begin{equation}
C(r)=\langle S_{j_\mathrm{imp1},\mathrm{l}}^z \> S_{j_\mathrm{imp1}+r,\mathrm{l}}^z \rangle
\end{equation}
and
\begin{equation}
D(r_\mathrm{imp})=\langle S_{j_\mathrm{imp1},\mathrm{l}}^z \> S_{j_\mathrm{imp2},\mathrm{l}}^z \rangle,
\end{equation}
where the distance $r_\mathrm{imp}=|j_\mathrm{imp1}-j_\mathrm{imp2}|$ corresponds to the distance between the two impurities.

We use the finite-size algorithm of the DMRG method to obtain the spin-spin correlation functions.  
At the beginning of the sweeping iteration, we introduce impurity sites into the pristine spin ladder obtained by the infinite algorithm. 
In this paper, we show the results of a 100-rung ladder system with two impurities for various $\alpha$.
We confirmed that the ground-state energy converges within a numerical error less than $10^{-10}J_1$ with the DMRG truncation number $m=230-250$. 

\section{Result}
\begin{figure}
\begin{center}
\includegraphics[scale=0.38]{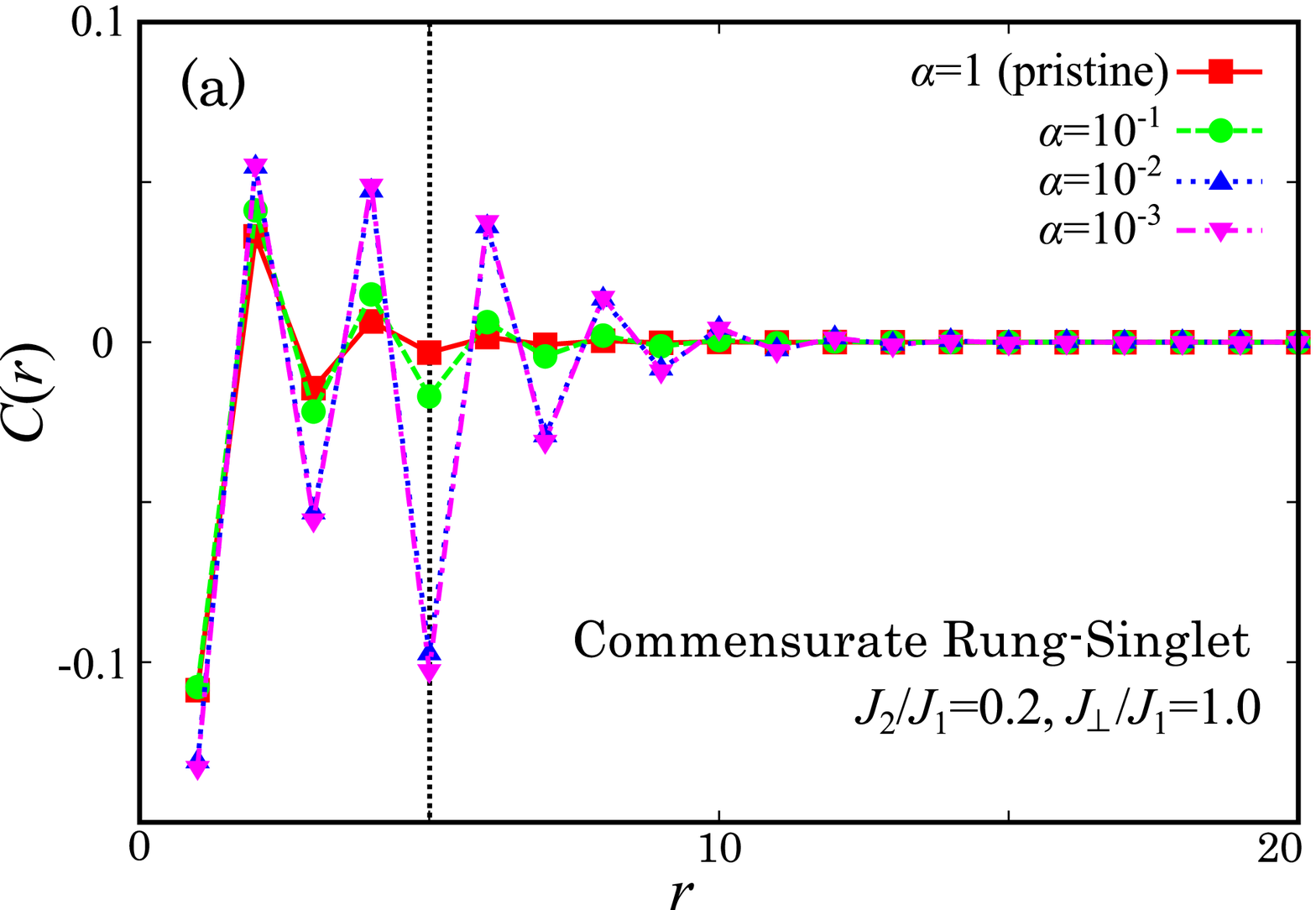}
\includegraphics[scale=0.38]{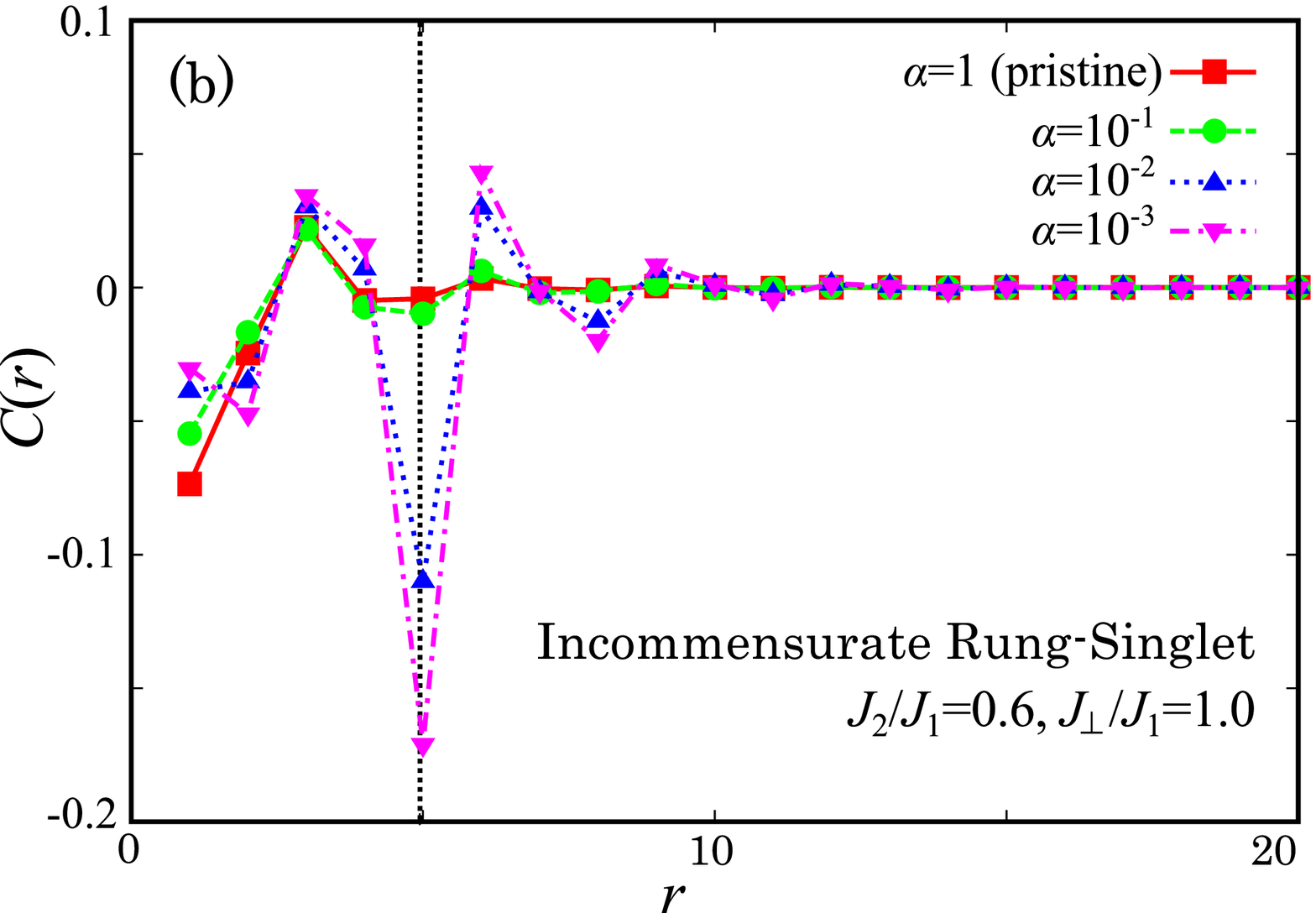}
\end{center}
\caption{(Color online) The spin-spin correlation function $C(r)$ in (a) the commensurate and (b) the incommensurate rung-singlet phases for various $\alpha$. One impurity is put on $r=0$ and the other one is located on $r=5$, indicated by dotted vertical line.}
\label{fig:cls}
\end{figure}

First, we demonstrate the enhancement of the spin-spin correlation around impurities.
The spin-spin correlation function $C(r)$ is calculated in both the commensurate and incommensurate rung-singlet phase for various $\alpha$.
Figure \ref{fig:cls} shows the $\alpha$ dependence of the spin-spin correlation function $C(r)$.
At $\alpha=1$, i.e., the pristine system, the wavenumber of the spin-spin correlation function is $\pi$ (non $\pi$) for commensurate (incommensurate) phase as seen in Fig.~\ref{fig:cls} (a) (Fig.~\ref{fig:cls} (b)).
We find the enhancement of $C(r)$ at $r=5$ corresponding to the distance of two impurities as decreasing $\alpha$ in both Fig.~\ref{fig:cls} (a) and (b).
Thus, investigating the spin-spin correlation between two impurities, $D(r_\mathrm{imp})$, is important to understand the effects of the impurity in our system.

\begin{figure}
\begin{center}
\includegraphics[scale=0.7]{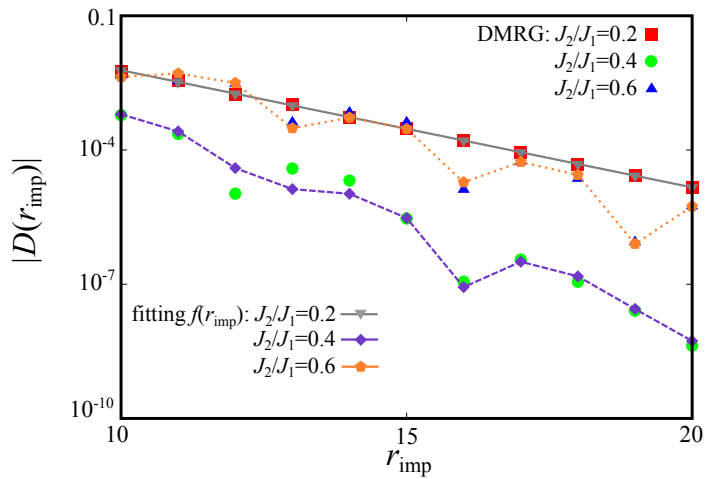}
\end{center}
\caption{(Color online) The absolute value of the spin-spin correlation function $|D(r_\mathrm{imp})|$ for various $J_2/J_1$ with fixed $\alpha=10^{-2}$ and $J_\perp/J_1=1$. The lines with points show the fitting function $f(r_\mathrm{imp})=A\exp(-r_\mathrm{imp}/\xi)|\cos(Q \, r_\mathrm{imp})|$. The parameters $(A,\xi,Q)$ are obtained by a nonlinear least squares method, the Marquardt-Levenberg algorithm.  The parameters $(A,\xi,Q)$ in the commensurate phase ($J_2/J_1=0.2$) obtained by fixing $Q=\pi$ are $(2.51\pm 0.15,1.66\pm 0.01, \pi)$. The parameters $(A,\xi,Q)$ in the incommensurate phase are $(34.4\pm 33.8,0.94\pm 0.06, 0.720\pi\pm 0.001\pi)$ for $J_2/J_1=0.4$ and $(38.0\pm 15.8,1.28\pm 0.04, 0.659\pi\pm 0.001\pi)$ for $J_2/J_1=0.6$.}
\label{fig:scil}
\end{figure}

We calculate $D(r_\mathrm{imp})$ by changing the distance of impurities $r_\mathrm{imp}$.
Figure \ref{fig:scil} denotes a log-linear plot of the absolute value of the spin-spin correlation function $|D(r_\mathrm{imp})|$.
$D(r_\mathrm{imp})$ in the commensurate phase, $J_2/J_1=0.2$, decreases exponentially without cosine-like oscillation because $|\cos(\pi \, r_\mathrm{imp})|=1$ for integer distance $r_\mathrm{imp}$.
On the other hand, $D(r_\mathrm{imp})$ in the incommensurate phase, $J_2/J_1=0.4$ and $J_2/J_1=0.6$, have a cosine-like oscillation with incommensurate wavenumber $Q\neq\pi$.
We obtain the incommensurate wavenumber by fitting a function $f(r_\mathrm{imp})=A\exp(-r_\mathrm{imp}/\xi)$ $\times|\cos(Q \, r_\mathrm{imp})|$ to our DMRG data. 
The obtained wavenumbers, $Q=0.720\pi\pm 0.001\pi$ for $J_2/J_1=0.4$ and $0.659\pi\pm 0.001\pi$ for $J_2/J_1=0.6$, are quite close to those obtained by the perturbation analysis, $\cos^{-1}(-J_1/4J_2)$, in the pristine system.
Therefore, we conclude that the wavenumber of the spin-spin correlation enhanced around impurities reflects the intrinsic wavenumber characterized by the frustration.

\section{Summary and discussion}
We have shown the spin-spin correlation functions in the frustrated two-leg spin ladder with non-magnetic impurities.
First, we demonstrate the enhancement of the spin-spin correlation around impurities by using DMRG method.
The behavior is identical to the effect of the impurity on the other quantum spin systems.
We also present the wavenumber of the spin-spin correlation enhanced by impurities.
The wavenumber reflects that of the spin-spin correlation in the pristine system with frustration.
The impurity-induced magnetic order reflects the intrinsic wavenumber of the spin-spin correlation in the singlet ground state.

The preceding experimental works of NMR and $\mu$SR showed that the AF-like magnetization is induced by the non-magnetic impurities.
Nevertheless, it remains controversial whether there is a long-ranged order or not.
In addition, since the two-leg spin ladder compound BiCu$_2$PO$_6$ has strong frustration, the impurity-induced magnetic order will be influenced by the incommensurate wavenumber like a spiral order.
These issues will be uncovered by INS measurements on Bi(Cu$_{1-x}$Zn$_x$)$_2$PO$_6$ in the near future.

\section*{Acknowledgements}
This work was partly supported by Grant-in-Aid for Scientific Research from MEXT (Grant No.24540387, No.24360036, No.23340093, and No.25287094) and by the inter-university cooperative research program of IMR, Tohoku University.
Numerical computation in this work was carried out on the supercomputers at JAEA.

\end{document}